\documentclass[nofootinbib,twocolumn,pre, showkeys]{revtex4}
\pdfoutput=1
\usepackage{dcolumn}
\usepackage{amsmath}
\usepackage{hyperref} % For all references
\usepackage{amsthm}
\newtheorem{remark}{Remark}

\usepackage{enumitem} % For enumerate in alphabetically
\usepackage{graphicx}
\usepackage{rotating}
\renewcommand{\vec}{\mathbf}
\usepackage{makecell}  % For newline in table cell

\begin{document}

 \title{On the initial value of PageRank}
\author {Krishanu Deyasi}
\email{krishanu.deyasi@iem.edu.in}
\affiliation{Department of Basic Science
\& Humanities,\\
Institute of
Engineering \& Management, Kolkata,\\
University of
Engineering \& Management, Kolkata\\
India}

\date{\today}

\begin{abstract}
 Google employs PageRank to rank web pages, determining the order in which search results are presented to users based on their queries. PageRank is primarily utilized for directed networks, although there are instances where it is also applied to undirected networks.
 In this paper, we have applied PageRank to undirected networks, showing that a vertex's PageRank relies on its initial value, often referred to as an intrinsic, non-network contribution. We have analytically proved that when the initial value of vertices is either proportional to their degrees or set to zero, the PageRank values of the vertices become directly proportional to their degrees. Simulated and empirical data are employed to bolster our research findings. Additionally, we have investigated the impact of initial values on PageRank localization.
\end{abstract}
%-----------------------------------------------------------------
% PACS Number (for show pacs \documentclass[twocolumn,pre,superscriptaddress, showpacs]{revtex4}
%\pacs{89.75.Hc}
%-----------------------------------------------------------------
\maketitle

\section{Introduction}
Complex network tools have been used in different fields from social, technological and  
biological networks \cite{newman2003structure,Albert02,Barabasibook,Newmanbook},
to analyze the structure and dynamics of the network. Various centrality metrics have been utilized to identify pivotal vertices within networks, including degree centrality, eigenvector centrality, Katz centrality, vertex betweenness centrality, edge betweenness centrality, and closeness centrality \cite{Newmanbook}, among others. Among these measures, the most straightforward is degree centrality, often simply referred to as degree. The degree of a vertex corresponds to the count of edges connected to that vertex.

PageRank stands out as another crucial metric utilized in network applications such as information retrieval, link prediction, and community detection \cite{Gleich15,chung09}.
In 1998, Brin and Page introduced the PageRank algorithm to rank web pages within their search engine \cite{BP98}. The PageRank score $x_i$, of a vertex $i$ within an undirected and unweighted network with $n$ vertices, is defined by the equation:

\begin{equation}
\label{equ1}
   x_i =  \alpha\sum_{j=1}^n A_{ij}\frac{x_j}{k_j}+(1-\alpha)\beta_i.
\end{equation}

Here, $0 < \alpha < 1$ represents the damping factor, also referred to as the teleportation factor (Mostly $\alpha 
= 0.85$~\cite{Langville11}). $A_{ij}$ denotes the $ij$th element of the adjacency matrix $\vec{A}$ corresponding to the network, while $k_j$ represents the degree of vertex $j$. $\beta_i$ stands for an initial value, often termed as the intrinsic, non-network contribution of vertex $i$. Extensive research has been conducted on the significance of $\alpha$ in the PageRank algorithm \cite{Langville04, Langville11, Pretto02, 
Fortunato07, Ding09}. But the significance of $\beta_i$ has not been explored \cite[p.~177]{Newmanbook}. In many cases the
value of $\beta_i$ has been taken as $1$ \cite{Langville04,Pretto02,Fortunato07}.

PageRank can be viewed as a random walk on a network \cite{fortunato2007}. Recently, there has been a surge of interest surrounding the concept of random walks on networks that incorporate stochastic resetting \cite{riascos2020random, christophorov2020, wald2021, huang2021random, ye2022random, chen2022random}. In stochastic resetting, the random walker is often reset from the initial vertex with a certain resetting probability whereas in PageRank each vertices are reset with uniform resetting probabilities \cite{riascos2020random, ye2022random}. Here, the initial value $\beta_i$ of PageRank represents the resetting probability of the vertex $i$. 

Researchers are always interested in optimizing PageRank \cite{Avrachenkov2006, De2008, Fercoq2012, Fercoq2014}.  Avrachenkov {\it et al.} \cite{Avrachenkov2006} theoretically showed that the PageRank of a web page can be increased by adding new edges from that web page. This idea was later generalized by Kerchove {\it et al.} \cite{De2008}. How PageRank can be optimizated using weighted edge has been analyzed by Fercoq {\it et al.} \cite{Fercoq2012, Fercoq2014}. It is expected that higher degree vertices get higher PageRank as a strong positive correlation is present between degree and PageRank 
\cite{Fortunato07}. A strong positive correlation between PageRank and degree signifies that the PageRank of a web page increases if that particular web page is connected to many web pages.  Here we examine whether a vertex having a low degree can come up with a higher PageRank by varying the initial value. 
%Apart from influence of degree Is it always true that irrespective of 
%any initial condition, vertices remain in the same position according to their PageRank? 
 %Does there exist any vertex which possess
%high PageRank instead of low degree? 
%Here we show that how initial value can play instrumental role on PageRank of a vertex. 

We have done another important study regarding the improvement of PageRank of a web page when connected to a high PageRank web page. This 
is similar to the analysis of whether the neighbor of a high PageRank web page also gets a 
higher PageRank value and consequently neighbors accumulate with high PageRank. 
Another way it can be thought of is whether the PageRank 
scores are evenly distributed over all the vertices in the network or most of the weight of the PageRank accumulates
in a small number of vertices. The latter case is known as localization.

Localization of the eigenvector is a common
phenomenon in many real-world networks and it is frequently suggested that the main cause of localization
is the presence of high degree vertices \cite{Goltsev12, Martin14}. We want to see whether the presence of high-degree vertices also propagates PageRank localization. Recently, empirical studies in PageRank on directed networks suggest that PageRank shows localization \cite{Ermann13}. Here we show that PageRank can be localized for some undirected networks also and the extent of localization of PageRank is proportional to the extent of localization of degree.

This article has been organized in the following way: 
in Section \ref{Aproof}, we show that if the initial value of the vertices is the
same or proportional to their  degrees, then
the PageRank remains the same or proportional to the degree of the vertices. We also show that if the initial value of the vertices is zero then PageRank
values become proportional to the degrees. In Section \ref{numerical}, our study on simulated and empirical
networks supports our findings.
In Section \ref{localization} we show that, for some networks, PageRank centrality can be localized 
and the value of $\beta_i$ can affect 
the localization of the PageRank centrality. Section \ref{conclusion} concludes with a discussion.

\section{Theoretical study}
\label{Aproof}
In the matrix form, the Equation (\ref{equ1}) can be written as:
\begin{equation}
\label{equ2}
       \vec{x} = \alpha \vec{AD}^{-1}\vec{x} + (1-\alpha)\pmb{\beta},   
\end{equation}
where $\vec{D}$ is the diagonal matrix with elements $D_{ii} = k_i$ and 
$\pmb\beta=(\beta_1,\beta_2,\dots,\beta_n)^T$.
Equation (\ref{equ2}) can be written as: 
\begin{equation}
 \label{equ3}
      \vec{x} = (\vec{I}-\alpha \vec{AD}^{-1})^{-1}(1-\alpha)\pmb{\beta}.
\end{equation}
 Since $|||\alpha \vec{AD}^{-1}|||_{\infty}=|\alpha||||\vec{AD}^{-1}|||_{\infty} = \alpha<1,$ Equation (\ref{equ3}) can be written as the sum of an infinite series \cite[p.~350]{horn2012matrix}:
 
 \begin{equation}
  \label{equ5}
  \vec{x} = \sum_{i=0}^\infty 
\alpha^i\left(\vec{A}\vec{D}^{-1}\right)^{i}(1-\alpha)\pmb{\beta}.
 \end{equation}

 Here  $\vec{AD^{-1}} = \vec{P}$ is a column stochastic matrix. Now we will show that if the network is undirected and connected, then $\vec{P}$ is diagonalizable.

 Note that $\vec{D^{-1/2}AD^{-1/2}}$ is a symmetric matrix, so there exists an invertible matrix $\vec{R}$ and a diagonal matrix $\vec{B}$ such that
  \begin{equation}
  \label{equ5a}
  \begin{split}
   \vec{D^{-1/2}AD^{-1/2}} & = \vec{RBR^{-1}}\\
   \vec{A} & = \vec{D^{1/2}RBR^{-1}D^{1/2}}
   \end{split}
 \end{equation}

 Now,
 \begin{equation}
  \label{equ5b}
  \begin{split}
  \vec{P} & = \vec{AD^{-1}}\\
          & = \vec{D^{1/2}RBR^{-1}D^{1/2}D^{-1}}~~~~~~~~~~~\text{[from Equation \ref{equ5a}]}\\
          & = \vec{D^{1/2}RBR^{-1}D^{-1/2}}\\
          & = \vec{D^{1/2}RB(D^{1/2}R)^{-1}}\\
          & = \vec{QBQ^{-1}},  ~~~~~~~~~~~\text{where $\vec{Q} = \vec{D^{1/2}R}$}
  \end{split}
 \end{equation}

 Hence the proof.

 Then  $(1-\alpha)\pmb{\beta}$ can be written as a linear combination of the
 eigenvectors $\vec{w}_j$ of $\vec{P}$, as
 
 \begin{equation}
 \label{equ6}
  (1-\alpha)\pmb{\beta} = \sum_{j=1}^{n} \vec{c}_j \vec{w}_j.
 \end{equation}

 Then the Equation (\ref{equ5}) reduces to
 \begin{equation}
 \label{equ7}
  \vec{x} = \sum_{i=0}^\infty \alpha^i \vec{P}^i \left(\sum_{j=1}^{n} \vec{c}_j \vec{w}_j\right).
 \end{equation}

 Let $1=\mu_1\ge\mu_2\ge \dots \ge\mu_n\ge -1$ be the eigenvalues of $\vec{P}$ \cite[p.~64]{van2011graph} corresponding to the
 eigenvectors $\vec{w}_1, \vec{w}_2, \dots ,\vec{w}_n$, respectively. Then
  \begin{equation}
  \label{equ8}
  \vec{x} = \sum_{i=0}^\infty \alpha^i \left(\sum_{j=1}^{n} \vec{c}_j \mu_j^i \vec{w}_j\right).
 \end{equation}
 After simplifying, we get
 \begin{equation}
 \label{equ9}
  \vec{x} = \sum_{j=1}^n c_j\left(\sum_{i=0}^\infty \left(\alpha\mu_j\right)^i\right)\vec{w}_j.
 \end{equation}
Because $|\alpha\mu_j| < 1~ \forall j=1,2,\dots, n$, we obtain
  \begin{equation}
\label{equ10}
  \begin{split}
    \vec{x} & = \sum_{j=1}^n c_j \frac{1}{1-\alpha\mu_j}\vec{w}_j.\\
   \end{split}
 \end{equation}

Therefore, the PageRank vector $\vec{x}$ of a network is the linear sum of eigenvectors of  $\vec{P}$. As $\vec{P}$ is a non-negative matrix so by the Perron–Frobenius theorem $\vec{w}_1$ will be the vector having entries non-negative entries \cite[p.~529]{horn2012matrix}. In particular, $\vec{w}_1$ is proportional to the degree of the network \cite[p.~63]{van2011graph}.

%%%%%%%%%%%%%%%%%%%%%%%%%% Remark 1 %%%%%%%%%%%%%%%%%%%%%%%%%%%%

\begin{remark}
\label{remark1}
If the value of $\pmb{\beta}$ is proportional to degrees, then Equation 
(\ref{equ6}) becomes:
\begin{equation}
 (1-\alpha)\pmb{\beta}=c_1\vec{w}_1.
\end{equation}
Therefore, Equation (\ref{equ10}) reduces to
\begin{equation}
 \vec{x} =  \frac{c_1}{1-\alpha}\vec{w}_1.
\end{equation}
Hence, PageRank centrality is proportional to the degree of the network.

\end{remark}
  
%%%%%%%%%%%%%%%%%%%%%%%%%% Remark 2 %%%%%%%%%%%%%%%%%%%%%%%%%%%%

\begin{remark}
\label{remark22}
 PageRank can also be quantified successively with an initial estimation of $\vec{x}$ from Equation (\ref{equ2}).
 Equation (\ref{equ2}) will be,
 \begin{equation} 
 \label{remark2}
  \vec{x}_1=\alpha \vec{A}\vec{D}^{-1}\vec{x}_0 + (1-\alpha)\pmb\beta,
 \end{equation}

where $\vec{x}_0$ is any non-negative vector (\cite[p.~170]{Newmanbook}).
If the initial value $\pmb\beta$ is zero, then from Equation (\ref{remark2}), we get
 \begin{equation}
  \vec{x}_1=\alpha\vec{A}\vec{D}^{-1}\vec{x}_0.
 \end{equation}
After $t$ iterative steps, we have
\begin{equation}
 \vec{x}_{t+1}=\left(\alpha\vec{P}\right)^t\vec{x}_0.
\end{equation}
Now, writing $\vec{x}_0$ as a linear combination of the eigenvectors $\vec{w}_j$ of $\vec{P}$, we get
\begin{equation}
 \vec{x}_0 = \sum_{i=1}^n c_i'\vec{w}_i,
\end{equation}
for some appropriate choice of $c_i'$. Then
\begin{equation}
 \vec{x}_{t+1}=\alpha^t \sum_{i=1}^n c_i'\mu_i^t\vec{w}_i.
\end{equation}
Since $|\mu_i|<1$ for all  $i\ne1$, we get $\vec{x}_{t+1}\to c_1'\alpha^t\vec{w}_1$ in the limit $t\to\infty$.
As $c'_1$ and $\alpha$ are constants and $\vec{w}_1$ is proportional to degree, it follows that PageRank
centrality is proportional to the degree centrality. 
\end{remark}

\section{Numerical study}
\label{numerical}
In the previous section, we have seen that the degree of a vertex plays an important role in its PageRank. So we take the initial PageRank value of a vertex as a function of its degree. Then we numerically examine four different cases depending on the degree and apply the same to various synthetic and real-world networks. In the first case, we initialize each vertex a value lesser than its degree by providing the reciprocal of the vertex degree, then we compute PageRank. In the second case, we give an equal non-zero value\footnote{For zero initial value, PageRank value is proportional to degree from Remark \ref{remark22}.} for all vertices, and then we compute PageRank.
The initial value is taken as proportional to the degree of the vertex in the third case. In the fourth case, the initial value of each
vertex is given a higher value by squaring the vertex degree. All of these four cases can be summarized as follows:
\begin{enumerate}%[label=(\roman*)]
\label{4conditioned}
 \item  $\beta_i$ is inversely proportional to the degree of vertex $i$.
 \item  All $\beta_i$'s are equal to one. 
 \item  $\beta_i$ is proportional to the degree of vertex $i$. 
 \item  $\beta_i$ is proportional to the square of the degree of vertex $i$.
\end{enumerate}

\begin{figure}
\begin{center}
  {\includegraphics[scale=0.65]{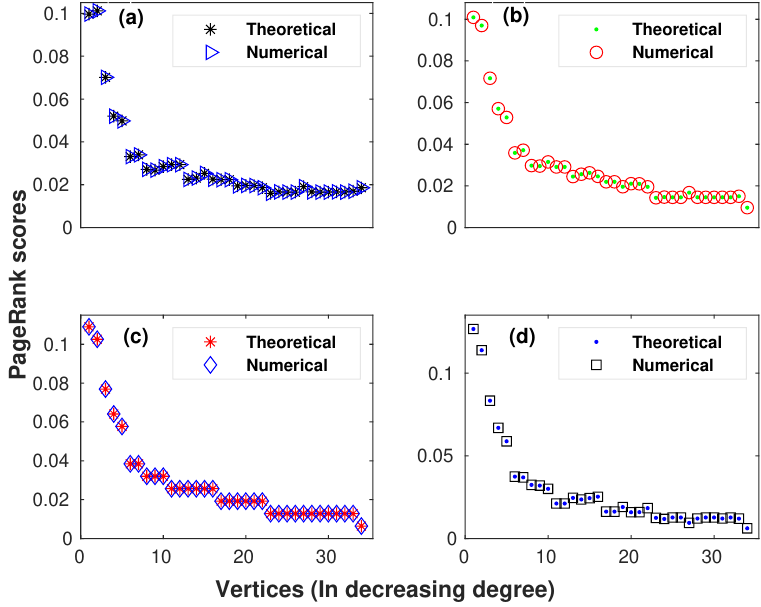}}
\end{center}
\caption{(Colour online) Theoretical versus numerical analysis on Zachary's karate club network \cite{zachary1977} for different initial values. When the initial value is
  (a) the inverse of the vertex degree, (b) equal for all vertices, (c) proportional to the vertex degree, 
  (d) proportional to the square of the vertex degree.}
  \label{fig:ThVsNm}
\end{figure}

First, we have conducted a numerical study using our analytical derivation of PageRank. In Equation (\ref{equ10}), we expressed the PageRank vector as a linear combination of the eigenvectors of the column stochastic matrix $\vec{P}$. Next, we have numerically compared the Equation (\ref{equ1}) with Equation (\ref{equ10}). For this analysis, we used Zachary's karate club network \cite{zachary1977}. Figure \ref{fig:ThVsNm} demonstrates that our analytical results perfectly match the numerical ones for all four initial values discussed above.
%%%%%%%%%%%%%%%%%%%%%%%%%%%% Theoretical vs Numerical Simulation %%%%%%%%%%%%%%%%%%%%%%%%%%%%%%%

%%%%%%%%%%%%%%%%%%%%%%%%%%%% ER network %%%%%%%%%%%%%%%%%%%%%%%%%%%%%%
\begin{figure}
\begin{center}
{\includegraphics[scale=0.47]{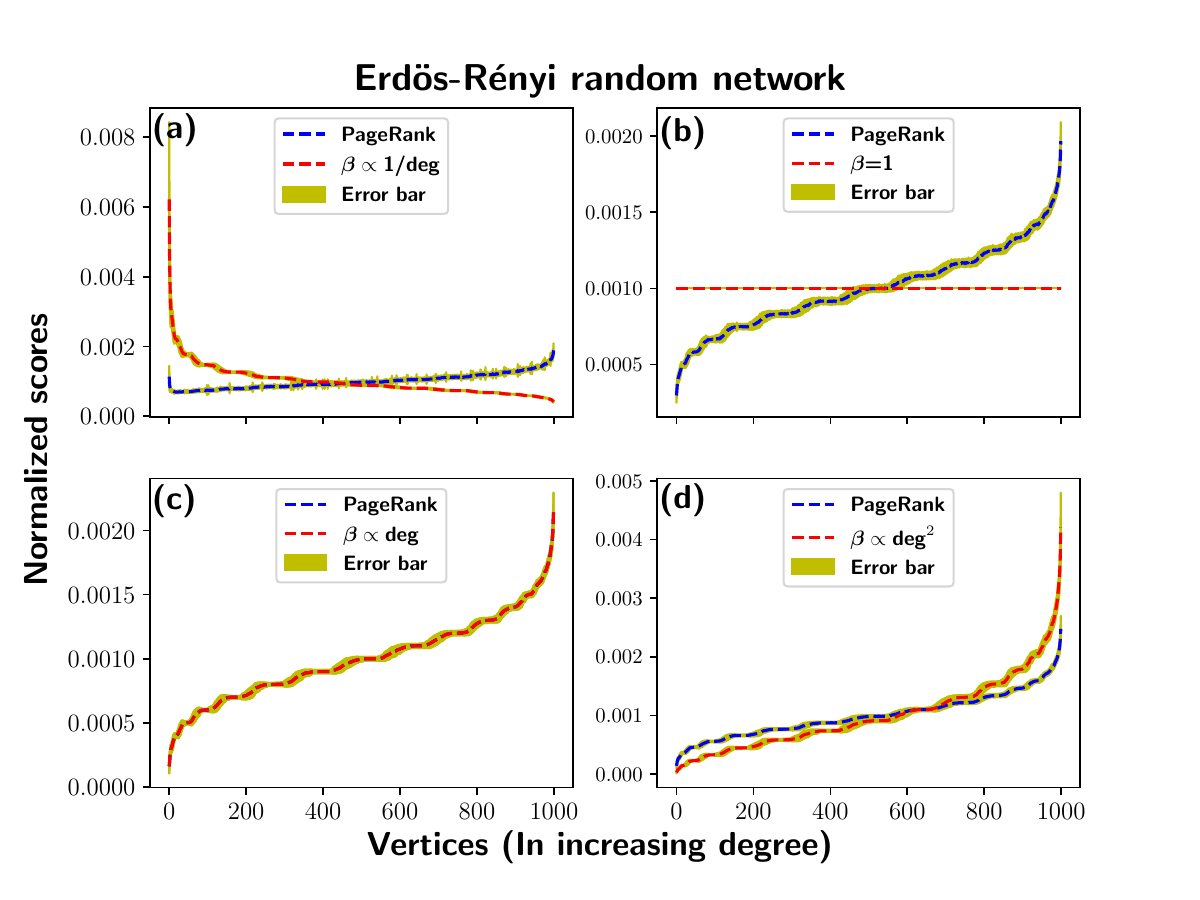}}
\end{center}
\caption{(Colour online) PageRank on Erd\"os-R\'eyni (ER) random network for different initial values. When the initial value is
 (a) inverse of the vertex degree, (b) equal for all vertices, (c) proportional to the vertex degree, (d) proportional to the square
 of the vertex degree.
 Random network model proposed by Erd\"os-R\'eyni \cite{Bollobas01}. Here the ER random network is generated with $1000$ vertices 
 and the probability of connecting two vertices is $0.01$. }
\label{fig:ER}
\end{figure}

%%%%%%%%%%%%%%%%%%%%%%%%%%%% WS network %%%%%%%%%%%%%%%%%%%%%%%%%%%%%%%
\begin{figure}
\begin{center}
  {\includegraphics[scale=0.47]{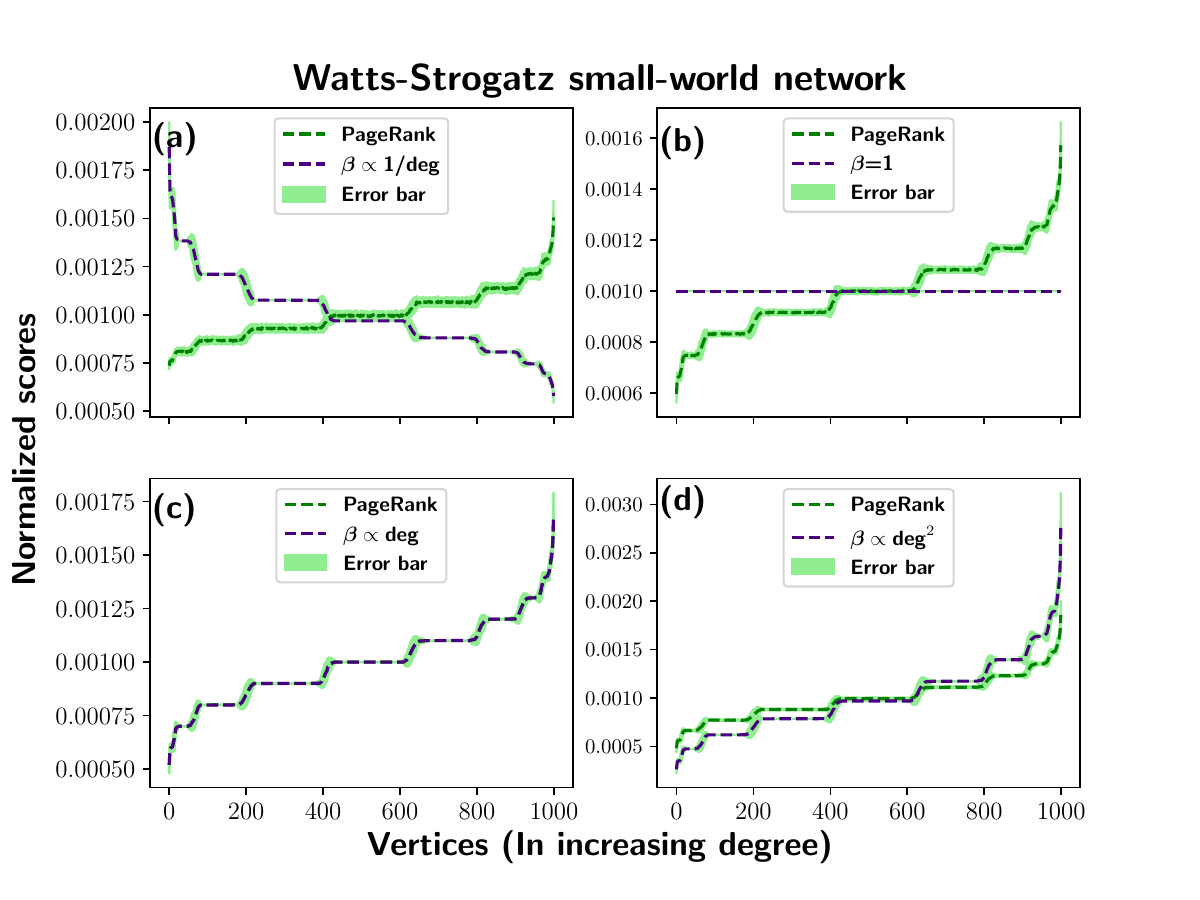}}
\end{center}
\caption{(Colour online) PageRank on the small-world network for different initial values. When the initial value is
  (a) the inverse of the vertex degree, (b) equal for all vertices, (c) proportional to the vertex degree, 
  (d) proportional to the square of the vertex degree.
   Small-world network proposed by Watts-Strogatz \cite{WS98}. Here the small-world network is generated by
   rewiring regular ring lattice of size $1000$
   and average degree $10$ with rewiring probability $0.4$.}
  \label{fig:WS}
\end{figure}

%%%%%%%%%%%%%%%%%%%%%%%%%%%  BA network %%%%%%%%%%%%%%%%%%%%%%%%%%%%%%%%%

\begin{figure}
\begin{center}
 {\includegraphics[scale=0.47]{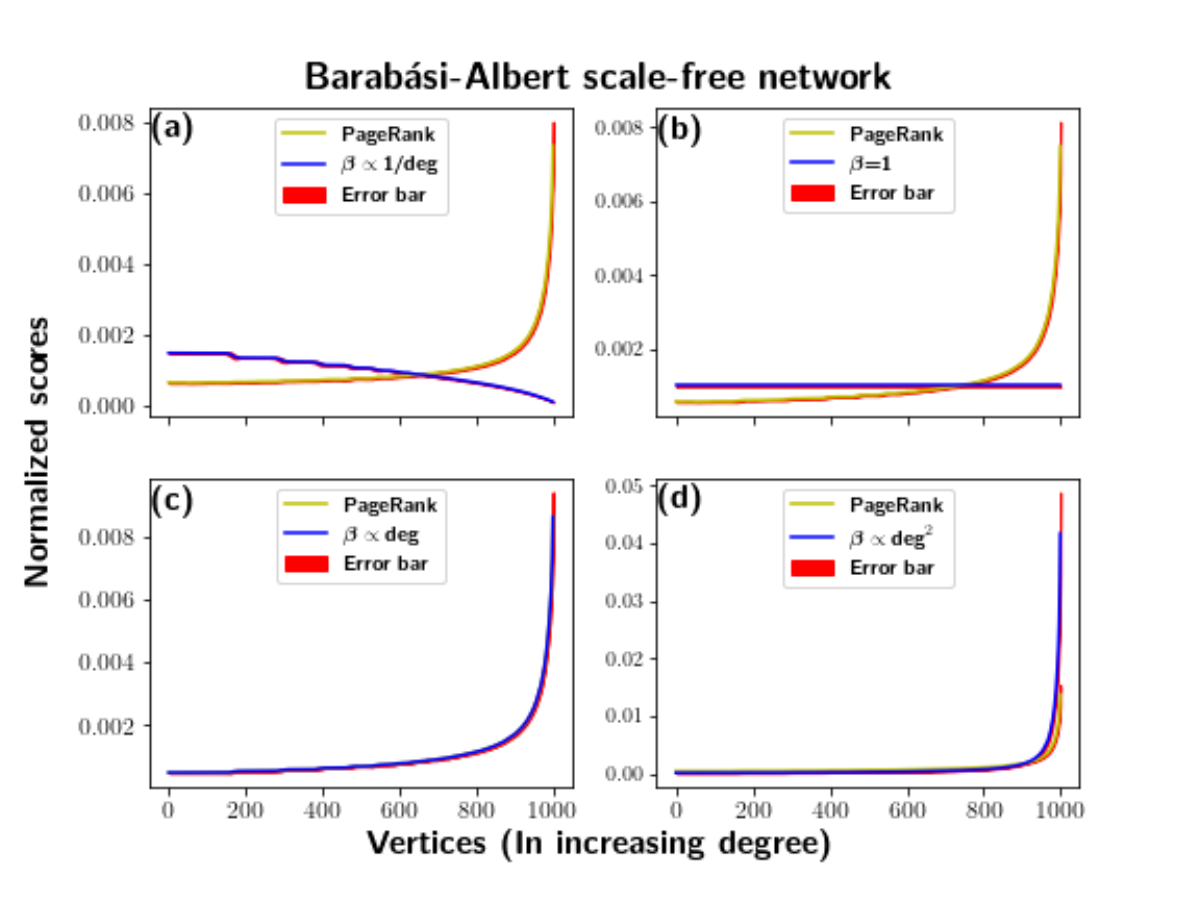}}
\end{center}
\caption{(Colour online) PageRank stability on the scale-free network for different initial values. When the initial value is (a) inverse of the vertex degree, (b) equal for all vertices,
  (c) proportional to the vertex degree, (d) proportional to the square of the vertex degree. Scale-free network proposed 
  by Barab\'asi-Albert~\cite{BA99}.
  Here the scale-free network is generated with size $1000$ and the size of the seed network is  $m_0=5$ and a new
  vertex is added with existing $m=5$ vertices.}
  \label{fig:BA}
\end{figure}

%%%%%%%%%%%%%%%%%%%%%%%%%%%  DD network %%%%%%%%%%%%%%%%%%%%%%%%%%%%%%%%%

\begin{figure}
\begin{center}
 {\includegraphics[scale=0.47]{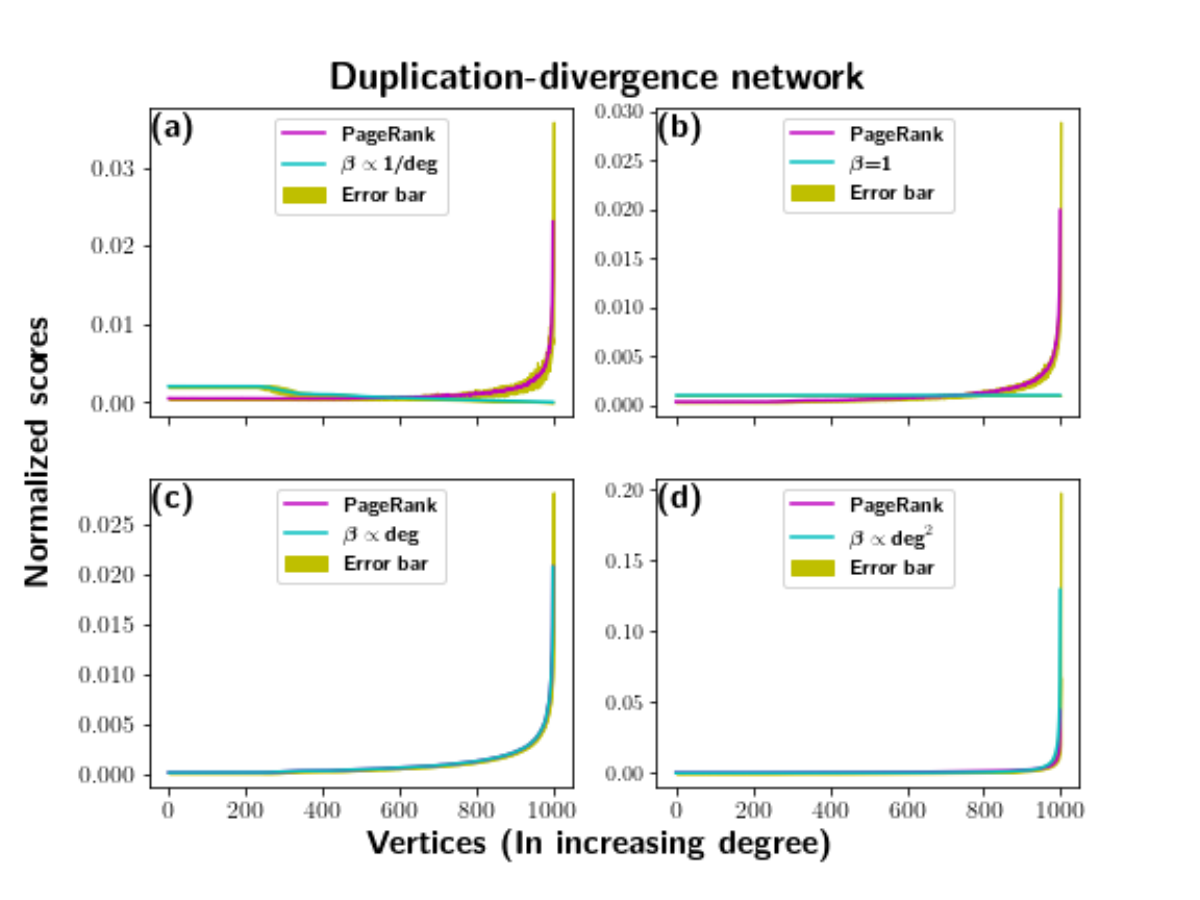}}
\end{center}
\caption{(Colour online) PageRank stability on the duplication-divergence network for different initial values. When the initial value is (a) inverse of the vertex degree, (b) equal for all vertices,
  (c) proportional to the vertex degree, (d) proportional to the square of the vertex degree. The duplication-divergence model proposed by I.~Ispolatov {\it et al.}~\cite{ispolatov2005duplication}.
  Here the duplication-divergence network is generated with size $1000$ with the link retention probability $\sigma = 0.4$.}
  \label{fig:DD}
\end{figure}
\subsection{Simulation on the model networks}
In our numerical simulation on the Erd\"os-R\'eyni (ER) random network \cite{Bollobas01}, we have observed that if
the initial value $\beta_i$ is proportional to the inverse of the degree of
vertex $i$ then PageRank score increases for a few numbers of low-degree vertices and for others PageRank 
is \emph{stable}. Here stable means modifications of $\beta_i$ of those vertices in the network do not cause dramatic changes in the
ranking of the vertices (Figure  \ref{fig:ER}(a)).
%Here  `stable' means, that the PageRank scores are positively correlated with the degree. 
If $\pmb\beta$ is proportional to the degrees of the vertices then 
PageRank centrality is proportional
to the degrees of the vertices (Figure  \ref{fig:ER}(c)). PageRank values are
stable when $\pmb\beta$ is either equal to one or proportional to the square of 
the degree of a vertex
(Figure \ref{fig:ER}(b) and Figure \ref{fig:ER}(d)). In comparison  with ER random
network, PageRank for the other three model networks that we have studied here, namely Watts-Strogatz (WS) small-world network \cite{WS98}, Barab\'asi-Albert (BA) scale-free networks \cite{BA99} and duplication-divergence network \cite{ispolatov2005duplication} are stable with any value of
$\pmb\beta$ that we have
considered (Figure \ref{fig:WS}, Figure \ref{fig:BA} and Figure \ref{fig:DD}). In the WS small-world network, PageRank values are monotonically increasing with respect to the degrees for all four different initial values (Figure \ref{fig:WS}). For all four initial values, PageRank for BA scale-free network and duplication-divergence network are gradually increasing with the increment of the degrees (Figure \ref{fig:BA} \& \ref{fig:DD}).
% In ER random network and 
% WS small-world network, step like PageRank score is because of the degree distribution
% pattern as both they show Gaussian degree distribution. And same degree vertices posses
% same PageRank. This step like with very small standard error suggests that PageRank score
% for a vertex is correlated with the degree irrespective their position in the network.
\\
%%%%%%%%%%%%%%%%%%%%%% Electric network %%%%%%%%%%%%%%%%%
\begin{figure}
\begin{center}
  {\includegraphics[scale=0.47]{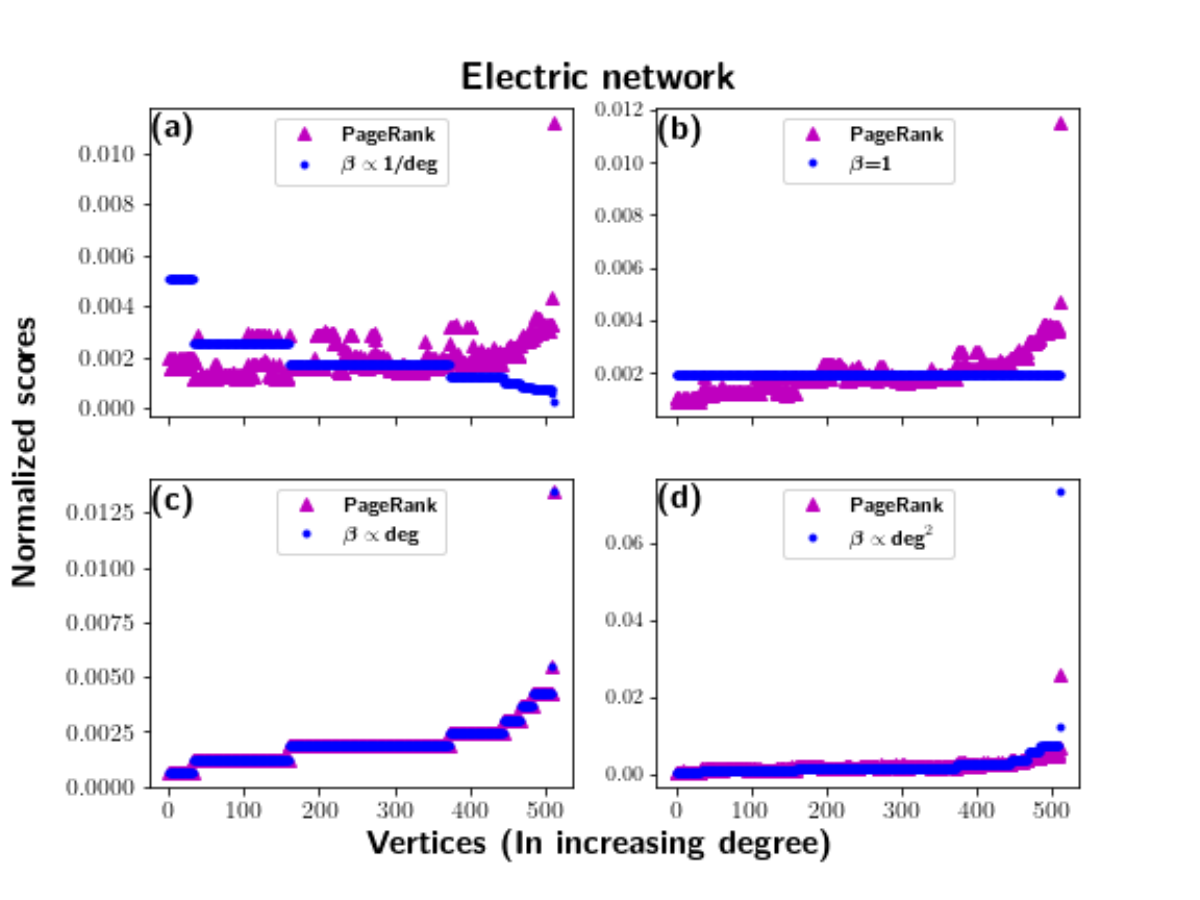}}
\end{center}
 \caption{(Colour online) PageRank on electric circuit network for different initial values. When the initial value is (a) inverse of the vertex degree, (b) equal for all vertices,
  (c) proportional to the vertex degree, (d) proportional to the square of the vertex degree. In the electric circuit network\cite{Emaildata},
  the vertices are
 power generating stations and edges
are the electric lines joining them.}
  \label{fig:Electric}
\end{figure}

%%%%%%%%%%%%%%%%%%%%%%%%%% Email network %%%%%%%%%%%%%%%%%%%%%%%%%

\begin{figure}
\begin{center}
  {\includegraphics[scale=0.47]{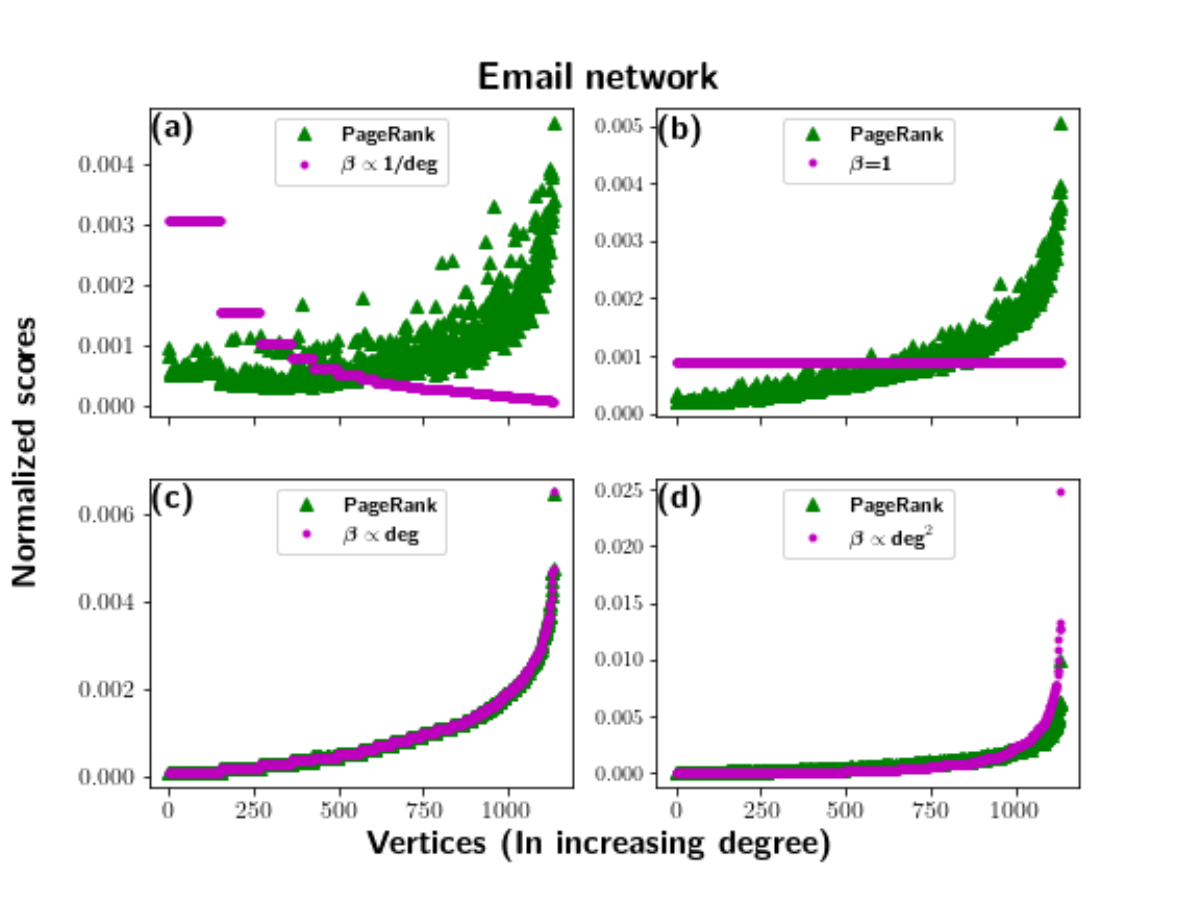}}
\end{center}
 \caption{(Colour online) PageRank on the email network for different initial values. When the initial value is (a) inverse of the vertex degree, (b) equal for all vertices,
  (c) proportional to the vertex degree, (d) proportional to the square of the vertex degree. In email network \cite{Emaildata}, 
  the vertices are 
  the users and two users are considered to be connected by an edge if one sends or receives email from the other.}
  \label{fig:Email}
\end{figure}

\subsection{Simulation on the real networks}
\begin{figure*}
\begin{center}
{\includegraphics[scale=0.6]{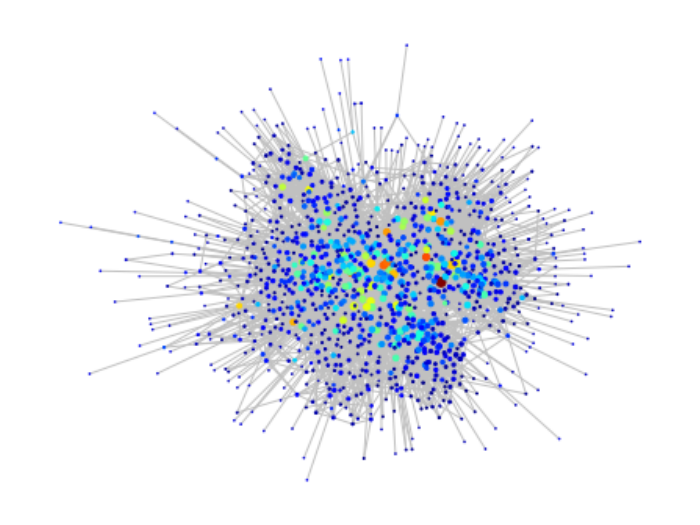}\includegraphics[scale=0.6]{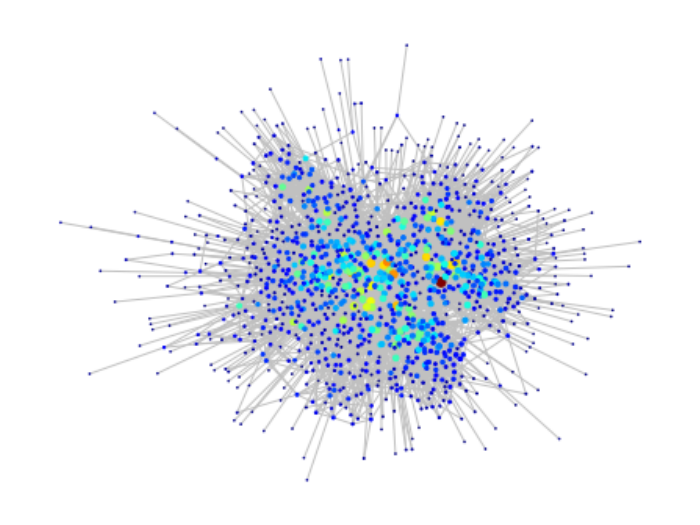}}\\
  \vspace*{-.3in}
{\includegraphics[scale=0.6]{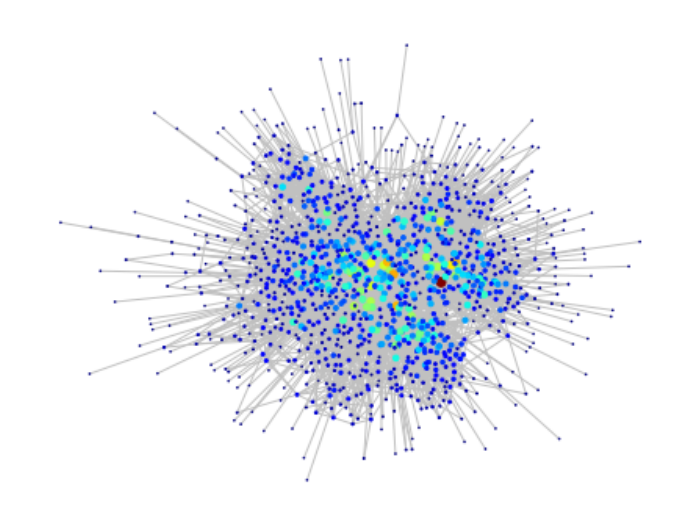}\includegraphics[scale=0.6]{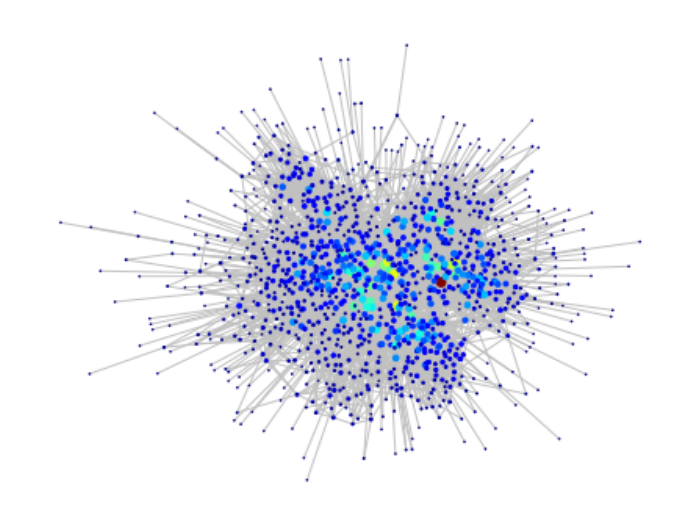}}\\
\vspace*{-2.15in} \hspace*{0.5in}  {\bf (a)} \hspace*{2.5in} {\bf (b)} \vspace*{2.25in}\\
 \vspace*{-.5in} \hspace*{0.5in}  {\bf (c)} \hspace*{2.5in} {\bf (d)} \vspace*{0.0in}
 
\end{center}
 \caption{(Colour online) PageRank on email network for different initial values. When the initial value is (a) inverse of the vertex degree, (b) equal for all vertices,
  (c) proportional to the vertex degree, (d) proportional to the square of the vertex degree. Vertex size is proportional to
  their degree, and color represent their PageRank where deep red color vertex represents high PageRank whereas
  deep blue represents low PageRank.}
  \label{fig:Email2}
\end{figure*}

For the electric circuit network, we have seen that when the initial value of the PageRank is inversely proportional to the degree of vertex then the PageRank of the vertices is not stable (Figure \ref{fig:Electric}(a)). In this first condition, we can see that some low-degree vertices are getting higher PageRank. For the second initial condition PageRank value is comparatively more stable than the previous case (Figure \ref{fig:Electric}(b)). For the third and fourth initial conditions, PageRank value highly correlated with the degree (Figure \ref{fig:Electric}(c) - (d)). In the email network, we have observed highly unstable PageRank values in accordance with the degree when the initial value of PageRank is inversely
proportional to the degree of vertex (Figure \ref{fig:Email}(a)).
When initial values are equal to one, we have seen a comparatively stable PageRank value than that in the
first case (Figure \ref{fig:Email}(b)).
PageRank value becomes proportional to the vertex degree when the initial value is proportional to the
degree (Figure \ref{fig:Email}(c)). When the initial value is proportional
to the square of the degree of vertex, higher degree vertices hold higher PageRank values
(Figure \ref{fig:Email}(d)). Figure \ref{fig:Email2} demonstrates these four different cases, where it
is observed that upon changing the initial value $\pmb\beta$ PageRank value changes. In this Figure, the warmer vertices have higher PageRank and the size of the vertices is proportional to the degree. We observe that Figure \ref{fig:Email2} (a) - (c) has a larger number of warmer vertices compared to \ref{fig:Email2} (d).
%%%%%%%%%%%%%%%%%%%%%%%%%%  Blogs network %%%%%%%%%%%%%%%%%%%%%%%%%
\begin{figure}
\begin{center}
 {\includegraphics[scale=0.47]{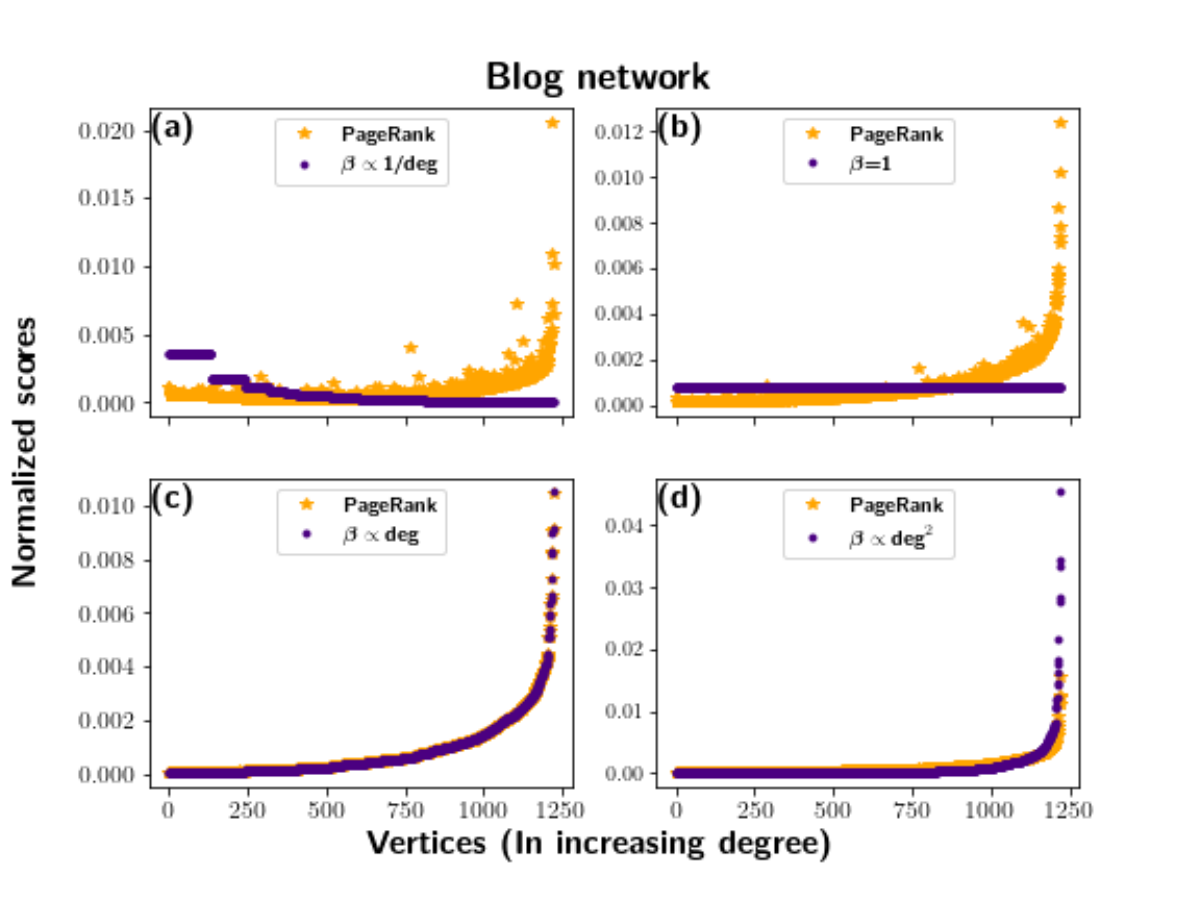}}
\end{center}
\caption{(Colour online) PageRank on blogs network for different initial values. When the initial value is (a) inverse of the vertex degree, (b) equal for all vertices,
  (c) proportional to the vertex degree, (d) proportional to the square of the vertex degree. In blog network \cite{Adamic05},
  vertices are the blogs and an edge 
  represents a hyperlink between two blogs.}
  \label{fig:Blog}
\end{figure}
In the blog network, we have observed that PageRank score 
fluctuates when the initial value $\pmb\beta$ of PageRank is 
inversely proportional to the degree of a vertex (Figure \ref{fig:Blog}(a)).
When all $\beta_i$ is equal to one, the PageRank value fluctuates but less than that in the previous case 
(Figure \ref{fig:Blog}(b)). When $\beta_i$ is proportional to the degree
of vertex $i$, PageRank score is proportional to $\pmb\beta$ (Figure
\ref{fig:Blog}(c)). PageRank values are stable when $\beta_i$ is proportional to 
the square of the degree
of vertex $i$ (Figure \ref{fig:Blog}(d)).
%%%%%%%%%%%%%%%%%%%%%%%%%% PGP network %%%%%%%%%%%%%%%%%%%%%%%%%
\begin{figure}
\begin{center}
 {\includegraphics[scale=0.47]{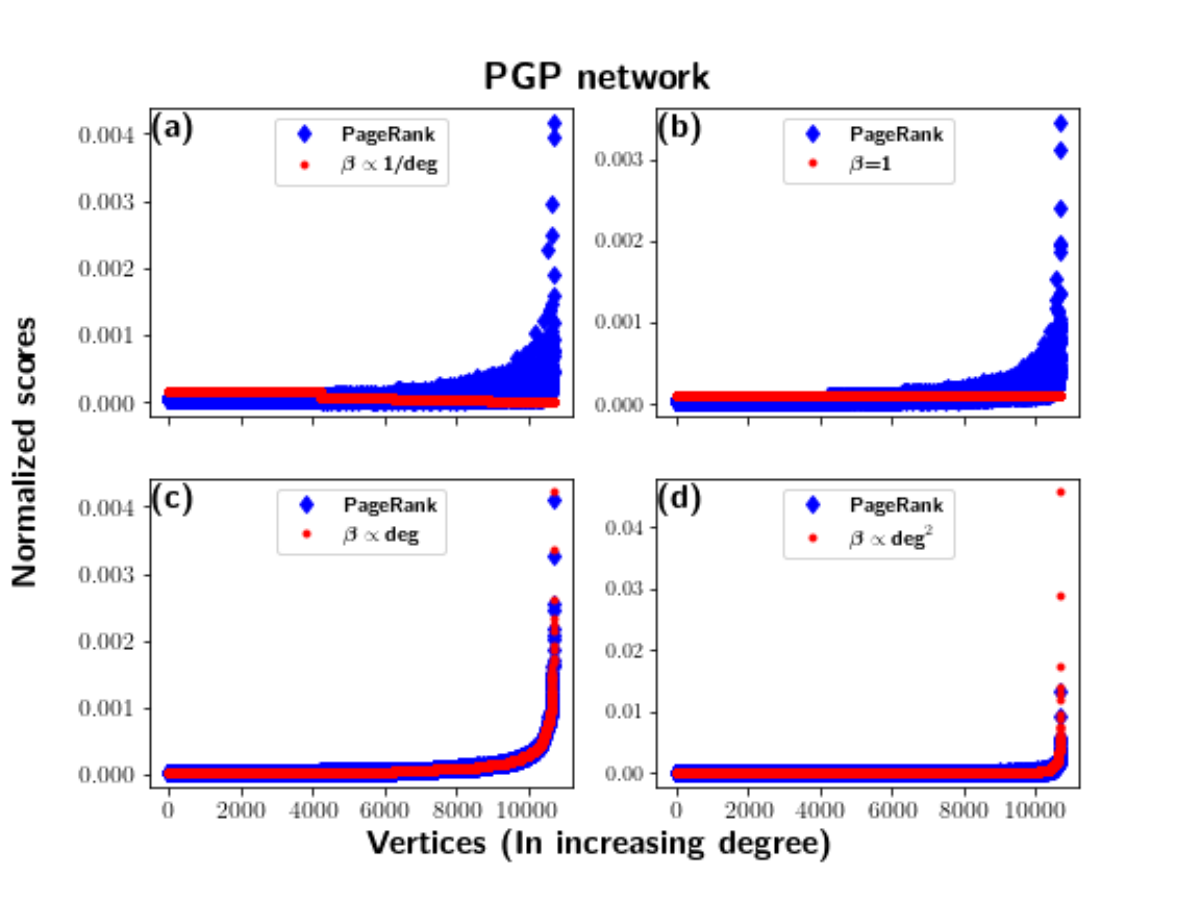}}
\end{center}
 \caption{(Colour online) PageRank on PGP network for different initial values. When the initial value is
(a) inverse of the vertex degree, (b) equal for all vertices,
  (c) proportional to the vertex degree, (d) proportional to the square of the vertex degree. In PGP network \cite{PGPdata},
  vertices are the user of Pretty Good
  Privacy (PGP) algorithm and edges are the interactions between them.}
  \label{fig:PGP}
\end{figure}
In Pretty Good Privacy (PGP) network, PageRank value of some high-degree vertices fluctuates
when $\beta_i$ is inversely proportional to the degree of a vertex $i$ (Figure \ref{fig:PGP}(a)).
A similar result is also observed when all $\beta_i$ is equal to one (Figure \ref{fig:PGP}(b)).
When $\beta_i$ is proportional to the degree of vertex $i$, PageRank value is 
proportional to the vertex degree (Figure \ref{fig:PGP}(c)). PageRank value increases with the increment
of the degree of vertices when $\beta_i$ is proportional to the square of the degree of
vertex $i$ (Figure \ref{fig:PGP}(d)).
%%%%%%%%%%%%%%%%%%%%%%%%%%  Gnutella peer-to-peer network %%%%%%%%%%%%%%%%%%%%%%%%%
\begin{figure}
\begin{center}
 {\includegraphics[scale=0.47]{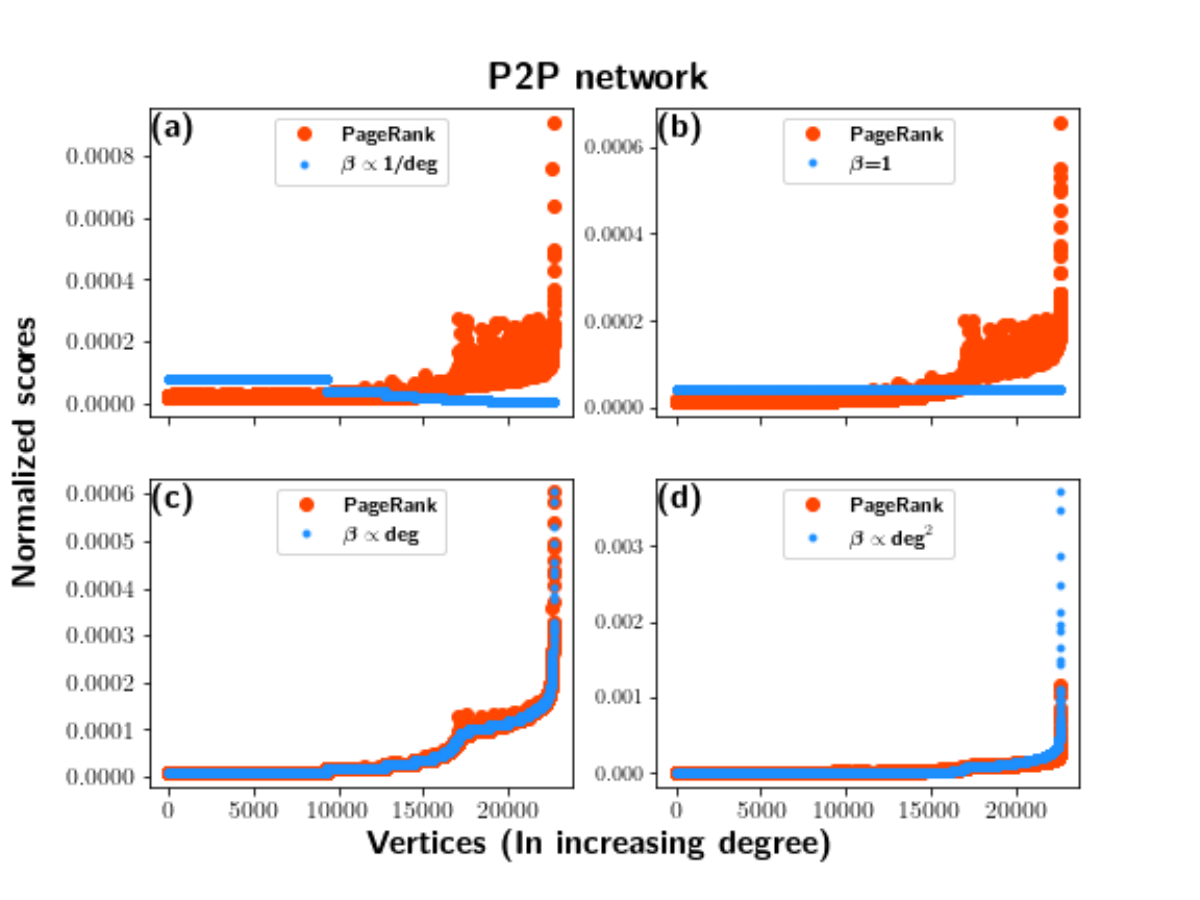}}
\end{center}
\caption{(Colour online) PageRank on Gnutella peer-to-peer network for different initial values. When the initial value is (a) inverse of the vertex degree, (b) equal for all vertices,
  (c) proportional to the vertex degree, (d) proportional to the square of the vertex degree. In Gnutella peer-to-peer 
  network \cite{Leskovec07}, the vertices are hosts in the
  Gnutella network topology and edges are the routes taken by data traveling between them.}
  \label{fig:P2P}
\end{figure}
In the peer-to-peer (P2P) network, we have seen a high fluctuation in PageRank value 
for some high degree vertices when $\pmb\beta$ is inversely proportional to the 
degree of a vertex 
(Figure \ref{fig:P2P}(a)). A similar result is also found when all $\beta_i$ is equal to one
(Figure \ref{fig:P2P}(b)). When $\beta_i$ is proportional
to the degree of vertex $i$, PageRank score is also proportional to the degree of vertex $i$ 
(Figure \ref{fig:P2P}(c)). PageRank score is stable when $\beta_i$ is proportional to
the square of the degree of vertex $i$ (Figure \ref{fig:P2P}(d)).
%%%%%%%%%%%%%%%%%%%%%%%%%% Internet network %%%%%%%%%%%%%%%%%%%%%%%%%
\begin{figure}
\begin{center}
 {\includegraphics[scale=0.47]{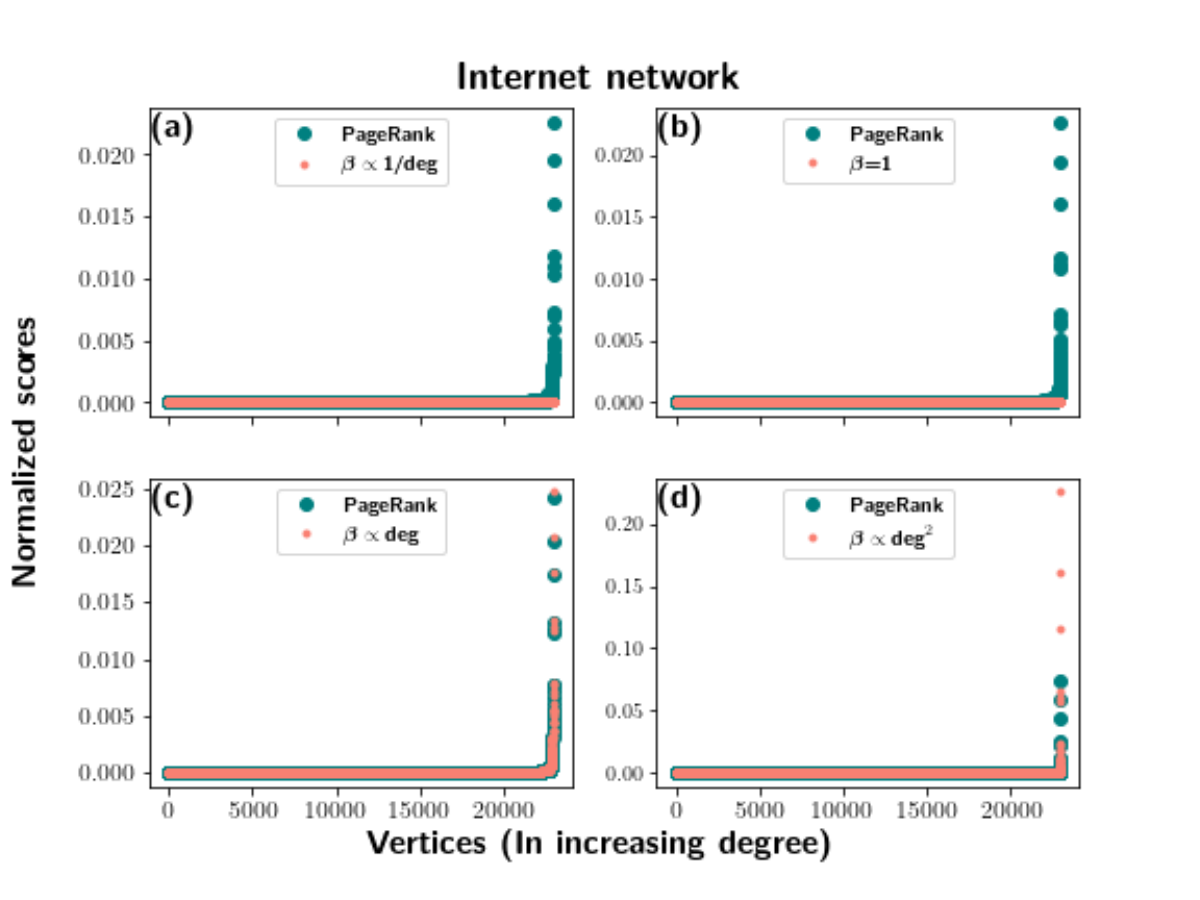}}
\end{center}
 \caption{(Colour online) PageRank on Internet network for different initial values. When the initial value is (a) inverse of the vertex degree, (b) equal for all vertices,
  (c) proportional to the vertex degree, (d) proportional to the square of the vertex degree. In Internet network \cite{Newman02},
  vertices are the
  autonomous systems (collection of computers and routers) and the edges show the route taken by the data
  traveling between them.}
  \label{fig:AS}
\end{figure}
Internet network is highly stable for all four different initial values of PageRank 
(Figure \ref{fig:AS}).

\begin{table*}[t]
\centering
\setlength{\tabcolsep}{4pt}
\begin{tabular}{llrrrrrrrrrr}
& {Network} & Vertices & $\rho_{\raisebox{-2pt}{\tiny 1}}~~$ & $\rho_{\raisebox{-2pt}{\tiny 2}}~~$ & $\rho_{\raisebox{-2pt}{\tiny 3}}~~$ & $\rho_{\raisebox{-2pt}{\tiny 4}}~~$ &  $IPR_{PR1}$ &  $IPR_{PR2}$ &  $IPR_{PR3}$ &  $IPR_{PR4}$ & $IPR_{Deg}$ \\ \hline
\begin{rotate}{90}
\hbox{\hspace{-2.95em}\textbf{Model}}
\end{rotate}
&  ER       &  1000  &    0.9465 & 0.9977 & 1 & 0.9977 & 0.0012 & 0.0012 & 0.0013 & 0.0015 & 0.0013 \\
&  WS       &  1000  &    0.9824 & 0.9973 & 1 & 0.9984 & 0.0010 & 0.0010 & 0.0011 & 0.0011 & 0.0011 \\
&  BA       &  1000  &    0.9984 & 0.9998 & 1 & 0.9881 & 0.0076 & 0.0078 & 0.0101 & 0.0226 & 0.0101 \\
&  DD       &  1000  &    0.9334 & 0.9840 & 1 & 0.9540 &  0.0685& 0.0441 & 0.0370 & 0.1405 & 0.0370 \\ \hline
\begin{rotate}{90}
\hbox{\hspace{-4.9em}\textbf{Real-world}}
\end{rotate}
& Electric  &  512   & 0.7218 & 0.9683   & 1 	  & 0.9575 & 0.0057 & 0.0061 & 0.0086 & 0.0484 & 0.0086\\
& Email     &  1133  & 0.9018 & 0.9870   & 1 	  & 0.9895 & 0.0036 & 0.0035 & 0.0045 & 0.0077 & 0.0045 \\
& Blog      & 1222   & 0.8634 & 0.9814   & 1 	  & 0.9886 & 0.0456 & 0.0148 & 0.0105 & 0.0202 & 0.0105 \\
& PGP       & 10680  & 0.6760 & 0.8108   & 0.9994 & 0.9298 & 0.0142 & 0.0076 & 0.0054 & 0.0337 & 0.0055 \\ 
& P2P       & 22663  & 0.9367 & 0.9835   & 0.9994 & 0.9866 & 0.0004 & 0.0002 & 0.0002 & 0.0007 & 0.0002 \\
& Internet  & 22963  & 0.9876 & 0.9921   & 0.9973 & 0.9286 & 0.1219 & 0.1197 & 0.1033 & 0.2577 & 0.1045 
\end{tabular}
\caption{$\rho_1, \rho_2, \rho_3, ~\text{and}~ \rho_4$ are the correlation coefficients between PageRank and degree when the initial value of the PageRank is inversely proportional to the degree of the vertex, equal to one, proportional to the degree of the vertex, proportional to the square of the degree of vertex, respectively. Theoretically, $\rho_3$ equals 1 regardless of the network type. However, in numerical computations with large network sizes, $\rho_3$ has not been observed to be exactly equal to 1 for PGP, P2P, and Internet networks, although it can often be approximated closely to 1. $IPR_{PR1}, IPR_{PR2}, IPR_{PR3}, IPR_{PR4}$'s are the inverse participation ratio of PageRank when the initial value of the PageRank is inversely proportional to the degree of vertex, equal to one, proportional to the degree of vertex, proportional to the square of the degree of vertex, respectively. $IPR_{Deg}$ is the inverse participation ratio of degree.}
\label{table}
\end{table*}
\section{Localization of PageRank centrality}
\label{localization}
Localization of a vector $\vec{v} = (v_1, v_2, \dots , v_n)^T$ means most of the weight of the vector $\vec{v}$ consists of only a few of the elements of the vector $\vec{v}$. On the other hand, if the weights of the entries of the vector $\vec{v}$ are evenly distributed over all the entries, then the vector $\vec{v}$  is called a delocalized vector.

The extent of localization can be measured by calculating the inverse participation ratio (IPR) \cite{Kawamoto16,Martin14}.
The IPR of a PageRank vector $\vec{v}$ is calculated as:
\begin{equation}
\label{ipr_equ}
 \text{IPR} = \frac{\sum_{i=1}^{n}v_i^4}{(\sum_{i=1}^n v_i^2)^2}.
\end{equation}

When IPR is of order $O(1)$ as $n\to\infty$, the PageRank vector $\vec{v}$ is localized. If IPR $\to 0$, $\vec{v}$
is delocalized \cite{Goltsev12, Kawamoto16}. IPR is mostly used in disordered mesocopic system to find the Anderson transition from localized phase to delocalized phase \cite{evers2008anderson}.
Although there does not exist any strict localization criteria for finite dimension system, but in numerical computation, if the IPR value close to 1 system seems to be localized \cite{Martin14}.
In the following section, we investigate whether PageRank exhibits localization in both simulated models and real-world networks. 
Additionally, we compute how the IPR value changes upon changing the initial values of PageRank.
%%%%%%%%%%%%%%%%%%%%%%%%%% Model localization %%%%%%%%%%%%%%%%%%%%%%%%%
\begin{figure}
\begin{center}
 {\includegraphics[scale=0.32]{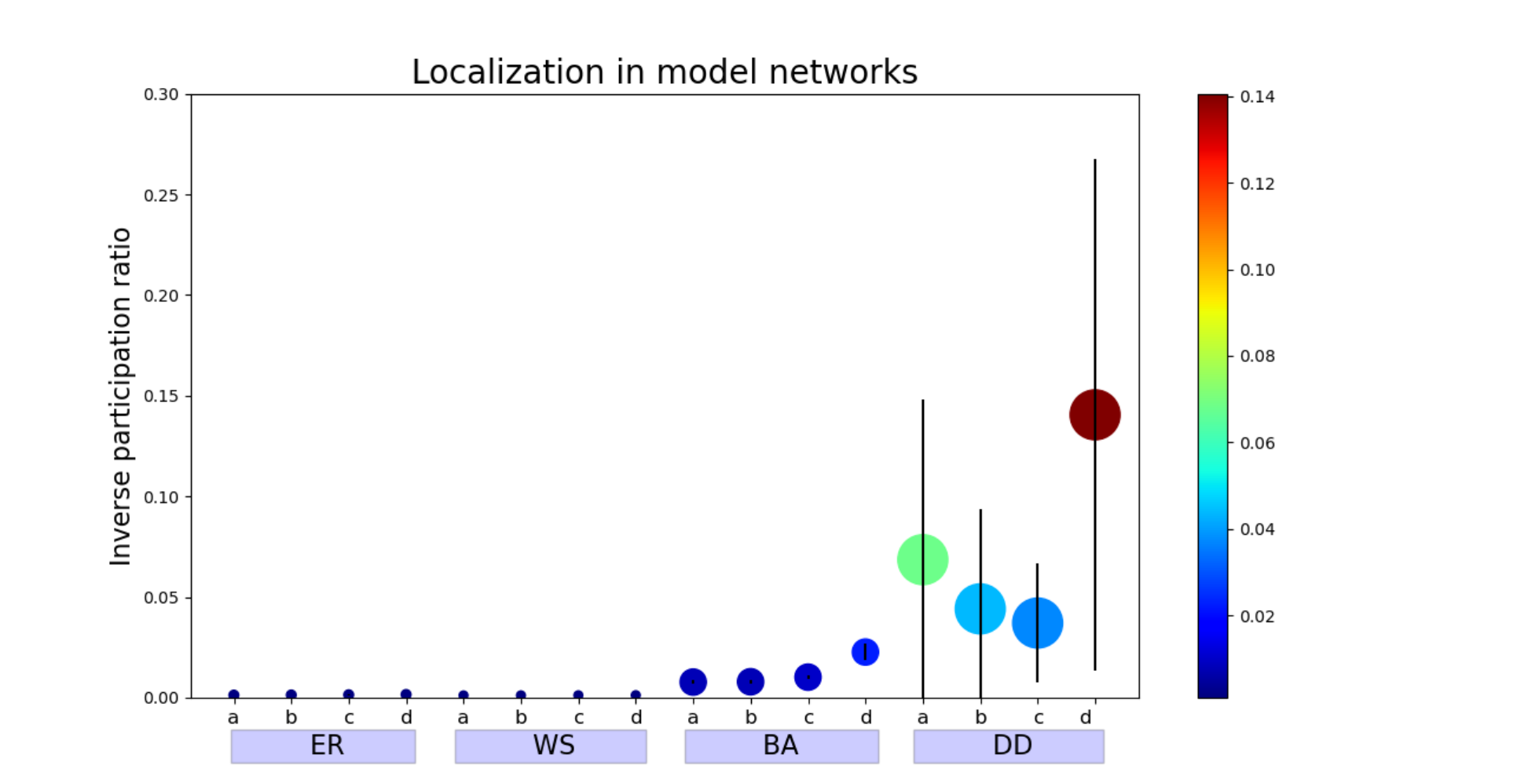}}
\end{center}
 \caption{(Colour online) Inverse participation ratio for the different initial values of PageRank for the four model networks. When the initial value is (a) proportional to the inverse of the vertex degree, (b)  equal to one,
  (c) proportional to the vertex degree, (d) proportional to the square of the vertex degree. The size of the circle is proportional to the inverse participation ratio value of the degree.}
  \label{fig:ML}
\end{figure}
%%%%%%%%%%%%%%%%%%%%%%%%%% DD localization %%%%%%%%%%%%%%%%%%%%%%%%%
\begin{figure}
\begin{center}
 {\includegraphics[scale=0.5]{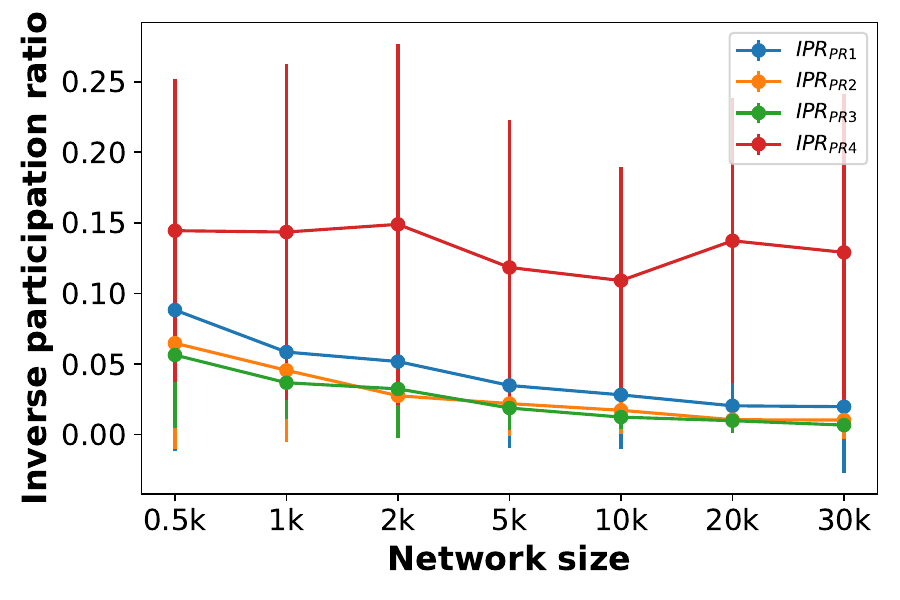}}
\end{center}
 \caption{(Colour online) Inverse participation ratio (IPR) for the different network sizes with different initial conditions of PageRank in duplication-divergence network. $IPR_{PR1}, IPR_{PR2}, IPR_{PR3}, IPR_{PR4}$'s are the inverse participation ratio of PageRank when the initial value of the PageRank is proportional to the degree of vertex, equal to one, proportional to the degree of vertex, proportional to the square of the degree of vertex, respectively.}
  \label{fig:DDL}
\end{figure}
\subsection{Localization of PageRank in model networks}
We observe that the PageRank of all the model networks that we have studied here (ER random network, WS small-world network, BA scale-free network and duplication-divergence network) does not show localization (Figure \ref{fig:ML}). In the ER random network and WS small-world networks, the lowest IPR value of PageRank is observed when the initial value of the PageRank is inversely proportional
to the degree of a vertex and the highest IPR value of PageRank is observed when the initial value of the PageRank is
square of the vertex degree (Table \ref{table}). In BA scale-free network the IPR value of PageRank is higher than both the ER random network and WS small-world network. The highest IPR value of PageRank is observed when the initial value is proportional to the square of the vertex degree and the lowest IPR value of PageRank is observed when the initial value is inversely proportional to the degree  (Table \ref{table} and Figure \ref{fig:ML}). In the duplication-divergence network, PageRank value seems to be localized when the initial value of PageRank is proportional to the square of the vertex degree (Figure \ref{fig:ML}). Due to the high error bars in the IPR values of PageRank on the duplication-divergence network, we conducted simulations across varying network sizes. Observations revealed a consistent decrease in the IPR values of PageRank across all four initial conditions as the network size increased (Figure \ref{fig:DDL}).
\\
%%%%%%%%%%%%%%%%%%%%%%%%%% Real localization %%%%%%%%%%%%%%%%%%%%%%%%%

\begin{figure*}
\begin{center}
 {\includegraphics[scale=.6]{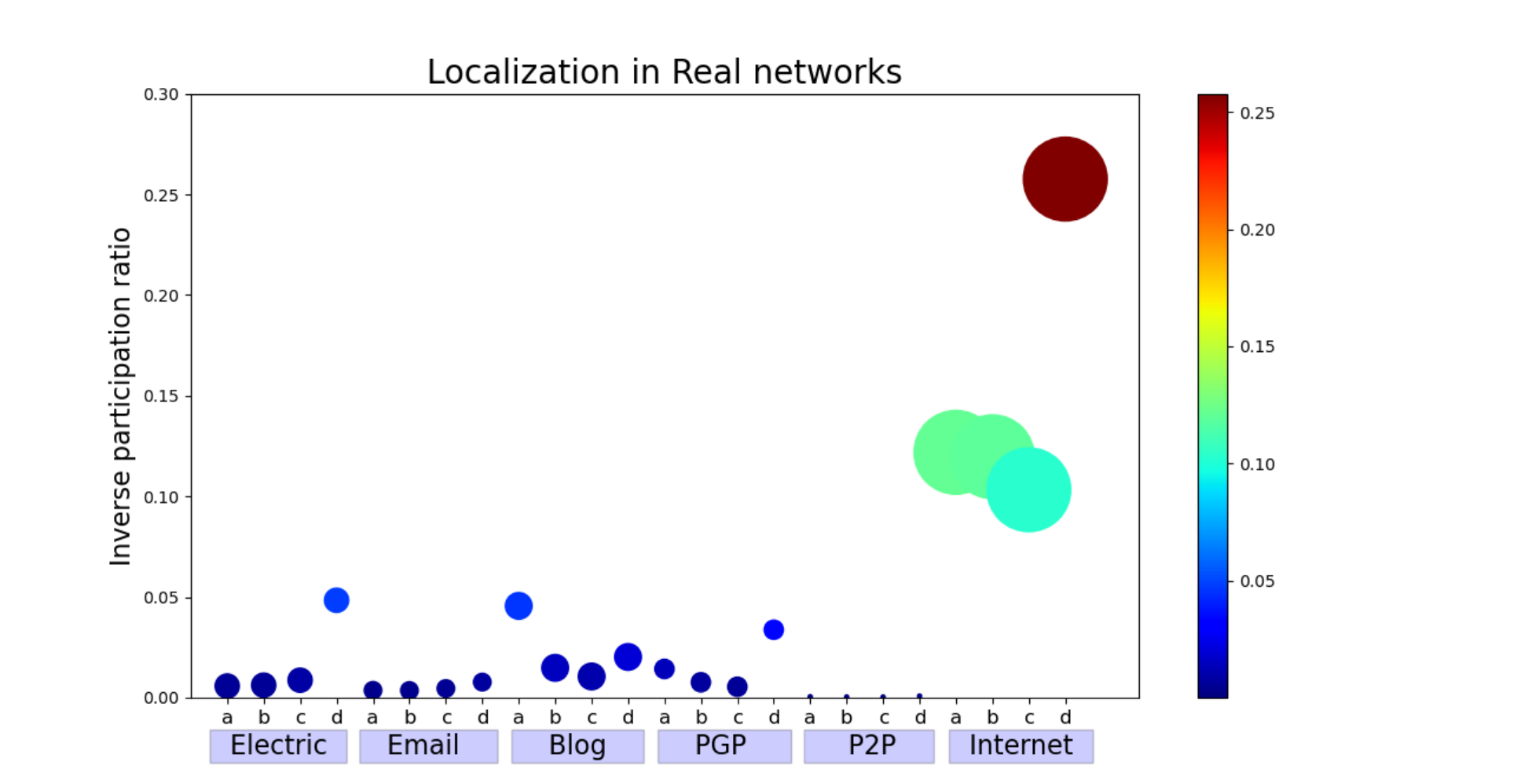}}
\end{center}
 \caption{(Colour online) Inverse participation ratio for different initial values of PageRank for the six real-world networks. When the initial value is (a) proportional to the inverse of the vertex degree, (b)  equal to one,
  (c) proportional to the vertex degree, (d) proportional to the square of the vertex degree. The size of the circle is proportional to the inverse participation ratio value of the degree.}
  \label{fig:RL}
\end{figure*}
\subsection{Localization of PageRank in real-world networks}
In the electric network, the PageRank is not localized for all four initial values of PageRank. The highest IPR value of PageRank is observed when the initial value of PageRank is proportional to the square of the vertex degree. The lowest IPR value is seen when the initial value is proportional to the inverse of the vertex degree (Table \ref{table} and Figure \ref{fig:RL}). The PageRank is also not localized in the email network for all four initial values of PageRank. The maximum and minimum IPR values of PageRank are observed when the initial value is proportional to the square of the vertex degree, equal to one, respectively (Table \ref{table} and Figure \ref{fig:RL}). In the blogs network, PageRank is not localized for all four initial conditions of PageRank. The highest IPR value of PageRank is observed when the value of $\pmb\beta$ is inversely proportional to the vertex degree. The lowest IPR value of PageRank is seen when $\pmb\beta$ is proportional to the degree 
of the vertex (Table \ref{table}). The PageRank vector is not localized in the PGP network. The highest IPR value of PageRank is observed
when $\pmb\beta$ is proportional to the square of the degree and the  lowest
IPR value of PageRank is seen when $\pmb\beta$ is proportional to the degree of the vertex (Table \ref{table} and Figure \ref{fig:RL}). The PageRank is not localized in the P2P network either. The highest IPR value of PageRank is observed when the initial value is
proportional to the square of the degree of a vertex and the lowest IPR is observed when the initial value of PageRank is either
equal to one or proportional to the vertex degree (Table \ref{table} and Figure \ref{fig:RL}). In the Internet network, the PageRank vector is localized. The highest IPR value of PageRank is observed when $\pmb\beta$ is proportional to the square of the vertex degree
and the lowest IPR value of PageRank is observed when $\pmb\beta$ is proportional to the degree of 
the vertex (Table \ref{table} and Figure \ref{fig:RL}).

\begin{figure}
\begin{center}
 {\includegraphics[scale=0.3]{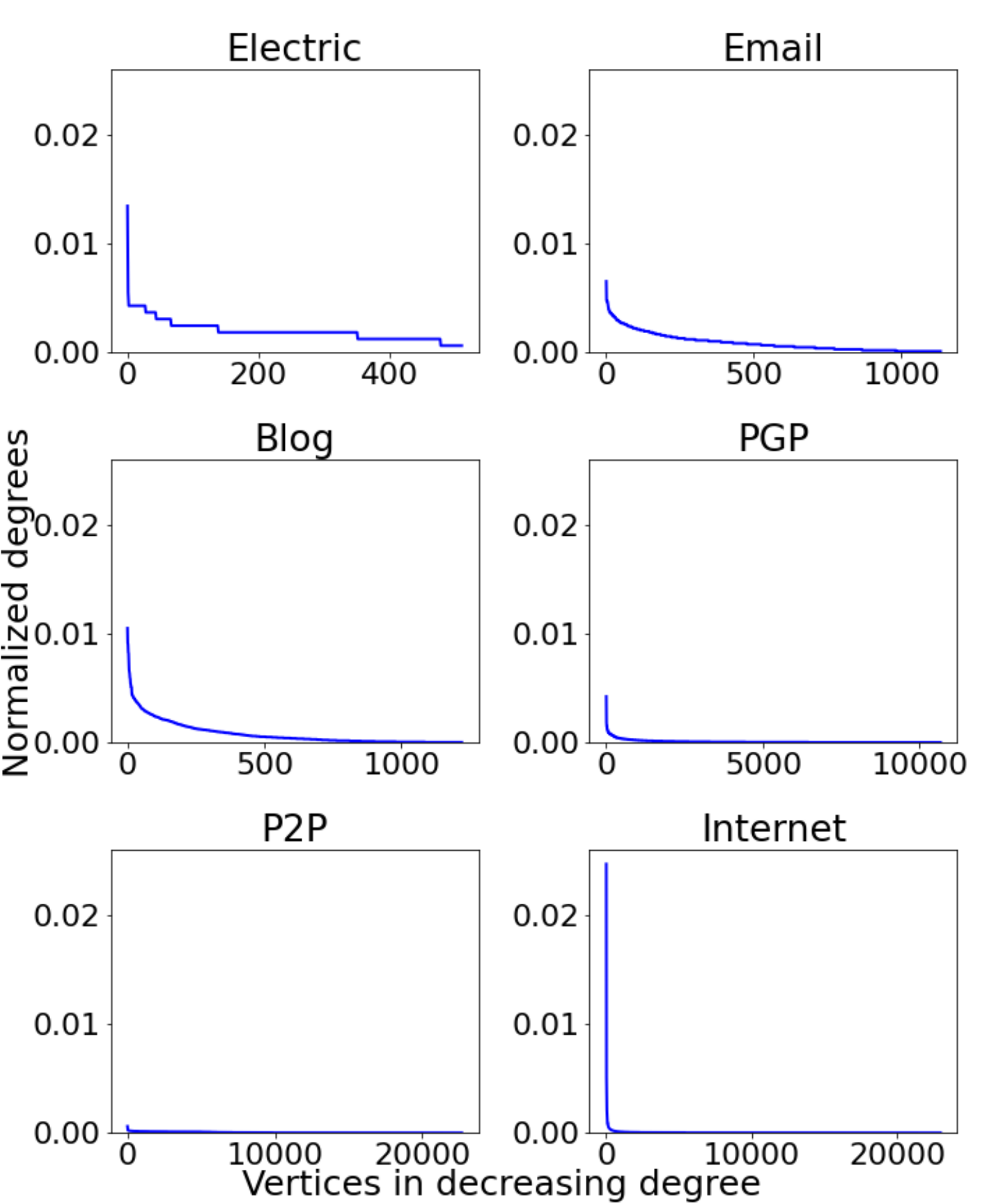}}
\end{center}
 \caption{Vertices degree of real-world networks.}
  \label{fig:RDD}
\end{figure}

\section{Discussion and conclusion}\label{conclusion}
In this work, we have shown that PageRank centrality depends on its initial value which is also called intrinsic, non-network 
contribution of the vertices. We have proved that if the initial value of PageRank is proportional to the degree of vertex then the PageRank score becomes proportional to
the degree of the vertex. Grolmusz \cite{grolmusz2015} made a similar conclusion. We have also shown that if the
initial value of PageRank is zero, then it also becomes
proportional to the vertex degree of the network. Our numerical simulation on the 
four model networks (ER random network, WS small-world network, BA scale-free network and duplication-divergence network) shows that the PageRank scores are more
resilient in WS small-world, BA scale-free and duplication-divergence networks than ER random network upon changing the initial value of PageRank. The instability of PageRank in the ER random network arises from its adherence to the Poisson distribution when the average degree of vertices is significantly less than the total number of nodes in the network. Otherwise, the network follows the binomial distribution \cite{Barabasibook, BA99}. Hence, the likelihood of encountering high-degree nodes in the ER random network is diminished \cite{BA99}. PageRank in the WS small-world network is stable. The rationale we have identified is that following the rewiring of edges in the regular ring lattice to form a small-world network, the degree of each node undergoes minimal change \cite{BA99}. Therefore, altering the initial value of PageRank does not influence the PageRank of the WS small-world network. PageRank in the BA scale-free network exhibits high stability thanks to the power law degree distribution, which increases the likelihood of encountering high-degree nodes \cite{BA99}. In this context, it is worth noting that PageRank remains stable in the scale-free network even when the network's edges are rewired \cite{ghoshal2011}. PageRank in the duplication-divergence network is also stable as the duplication-divergence network also follows power-law degree distribution for the retention probability $\sigma=0.4$ \cite{ispolatov2005duplication}.
Among all real-world networks studied 
here,  electric, email, blog, PGP, and P2P networks, PageRank values are unstable when the initial value of PageRank is either inversely proportional to
the vertex degree or equal to one. This is because, across all these networks, the likelihood of encountering high-degree nodes is low (Figure \ref{fig:RDD}). On the other hand, PageRank on the Internet
network is highly stable for all the initial values $\pmb\beta$ because the chances of getting the high degree are high (Figure \ref{fig:RDD}).

We have shown that the PageRank centrality can show localization
 and the extent of localization depends on the initial value. None of the four synthetic networks we have examined exhibit PageRank localization. Only the Internet network, among the six real-world networks, demonstrates PageRank localization.  In synthetic networks,
we have seen that the IPR value of PageRank increases with the increment of 
$\pmb\beta$ for ER random network, WS small-world network and BA scale-free network. We did not observe such a correlation in the duplication-divergence network or any of the real-world networks examined in this study. 
As there are high correlations between degree and PageRank in all synthetic and real networks, consequently the extent of localization for degree and PageRank of all the networks have similar magnitude (Table~\ref{table}). The localization of PageRank in the Internet network can be attributed to the existence of hub nodes, which are vertices with exceptionally high degrees (Figure \ref{fig:RDD}). In contrast, the other real-world networks we have examined lack any hub nodes (Figure \ref{fig:RDD}).  Thus, the presence of a hub vertex in a network fosters the localization of PageRank, rather than impacting its initial value.

In conclusion, we see that the initial value of PageRank can influence the PageRank score, and consequently the ranking of the web page. So specific and careful choice of initial value may enhance PageRank. The initial value of a web page can be given based on the textual relevance of the search query \cite[p.~177]{Newmanbook}. Accordingly, one can push up PageRank of their web page by giving more relevant words on their web page. This could yield favorable results for web pages with low degrees, as there's no influence on the ranking of highly ranked pages. The initial value of PageRank may slightly affect the inverse participation ratio, but it lacks the ability to influence the localization in PageRank centrality. Localization in PageRank centrality occurs when the network contains hub nodes.

\section *{{\bf Network resources:}}

{\bf Four simulated networks:} Erd\"os-R\'enyi (ER) random network is constructed with 1000 vertices and
two vertices are connected with probability $p = 0.01$ \cite{Bollobas01}.
In Watts-Strogatz (WS) small-world network,  the number of vertices is $1000$, average degree of initial
regular graph is $10$ and the rewiring probability is $0.4$ \cite{WS98}.
Barab\'asi-Albert (BA) scale-free network is generated with $1000$ vertices and size of the seed network
$m_0=10$. A new vertex is added with the existing $m=10$ vertices \cite{BA99}. The duplication-divergence network has been generated with 1000 vertices with the link retention probability $\sigma = 0.4$ \cite{ispolatov2005duplication}

{\bf Electric circuit network:} The vertices in the electric circuit network are power generating stations and edges are the electric lines joining the generating stations. The data were downloaded on 31 August 2016 from \cite{uri} and used in \cite{Martin14}.

{\bf E-mail network:} E-mail network is generated based on the exchange of emails between the members
of the University of Rovira i Virgili (Tarragone).
Here the vertices are the users, and two users are connected by an edge if one has sent or received email
from others. The number of vertices
and edges are 1133 and 5451, respectively. The data were downloaded on 31 August 2016 from \cite{alex} and used in \cite{Emaildata}.

{\bf Blogs network:} In the blogs network, vertices are the blogs and an edge represents a hyperlink between two blogs.
The number of vertices and edges are 1222 and 16714, respectively.
The data were downloaded on 31 August 2016 from KONECT \cite{konect} and used in \cite{Adamic05}.

{\bf Pretty Good Privacy network:} In the Pretty Good Privacy network, the vertices are the users of the Pretty Good Privacy (PGP) algorithm 
and edges are the interactions between them. The number of vertices
and edges are 10680 and 24316, respectively. The data were downloaded on 31 August 2016 from \cite{alex} and used in \cite{PGPdata}.

{\bf Gnutella peer-to-peer (P2P) network:} In the peer-to-peer network Gnutella, nodes are individual computers and edges are connections between them.
The number of vertices is 22663 and the number of edges is 54693. P2P network data were downloaded  on 31 August 2016 from 
SNAP \cite{SNAP} and used in \cite{Leskovec07, Ripeanu02}.

{\bf Internet network:} The vertices in the Internet network are autonomous systems (collection of computers and
routers \cite{Newmanbook}) and the edges 
are the routes taken by the data traveling between them. The numbers of vertices and edges are 22963 and
48436, respectively. Internet data were downloaded on 31 August 2016 from \cite{Newmanwebsite} 
and used in \cite{Newman02}.

\vspace{0.1in} 
We have considered the underlying undirected structure of all the networks. The giant component 
(i.e., the largest connected sub-network) has been considered when the network is not connected.

\section{Declarations}
  \subsection*{Competing interests:} The author declares no conflict of interests of a financial or personal nature. 
  
  \subsection*{Funding} No funding was received during this study.
  
  \subsection*{Availability of data and materials} All the data used in this study are freely available. The proper references and data sources have been mentioned in the text.

\begin{acknowledgments}
  The author thanks Anirban Banerjee and Abhijit Chakraborty for critical reading and suggestions.
  Author also thanks Prof. Asok Kumar Nanda for helping to prepare the manuscript. Author also thanks Travis Martin for FIG. 5 construction. 
   Author gratefully acknowledges the financial support from CSIR, India through a doctoral fellowship for an early version of this paper. 
\end{acknowledgments}

% References
\bibliography{PageRank}

\end{document}